\documentclass[3p,times]{article}

\usepackage{amsmath}
\usepackage{amsthm}
\usepackage{booktabs}

\usepackage{amssymb}

\usepackage[figuresright]{rotating}
\usepackage{tcolorbox}

\usepackage[font=footnotesize,labelfont=bf]{caption}
\usepackage[font=footnotesize,labelfont=bf]{subcaption}

\usepackage[english]{babel}
\title{A Note on the Evolution of Covid-19 in Italy}

\author{Giuseppe Dattoli, Emanuele Di Palma,\\ Silvia Licciardi\footnote{Corresponding author: silviakant@gmail.com, silvia.licciardi@enea.it, orcid 0000-0003-4564-8866, tel. nr: +39 06-94005421.}, Elio Sabia  \\[1.3ex]
	ENEA - Frascati Research Center, Via Enrico Fermi 45, 00044, \\Frascati, Rome, Italy  
}

\date{\today}
\begin{document}

	\maketitle

	\begin{abstract}
	We employ methods largely exploited in Physics, in the analysis of the evolution of dynamical systems, to study the pattern of the Covid-19 infection in Italy. The techniques we employ are based on the use of logistic function and of its derivative, namely the Hubbert function. The latter is exploited to give a prediction on the number of infected per day. We also mention the possibility of taking advantage from other mathematical tools based e.g. on the Gompertz equation and make some comparison on the different predictive capabilities.
	\end{abstract}
	
	\section{Introduction}	

The Covid-19 Pandemia, which is worldwide raging, does not allow the deve\-lopment of multiple strategic defensive scenarios. Viruses are entering the human body, cause an infection and there is no available anti-virus drugs to contrast Covid-19.\\

The restriction of social contacts is therefore the only (hopefully) efficient mean to slow down the diffusion of the infectious disease, thus allowing the sanitary structures to operate without the pressure of an emergency stress.\\

We have therefore asked ourselves whether tools, borrowed from other di\-sciplines, may be exploited as a predictive mean to understand how the number of Covid-19 infections increases with time, what is the maximum number of expected infections and when the rate of infections is expected to decrease.\\

In recent times the point of view, we are going to propose has been exploited for the understanding of bacterial growth \cite{Allen} or of other mechanisms mediated  by bacterial growth, such as global biogeochemical cycles \cite{Fenchel}, human gut health \cite{Flint}, wastewater treatment \cite{Sevious} or in totally different fields like cancer growth and metastsys dissemination \cite{DGDO} and in the physics of Free Elecron Laser \cite{DFel} as well.\\

In the following we will make use of the logistic function \cite{Cramer} and of its variants \cite{Tjorve}, to get a reasonable fit of the data provided by the Italian Department of health\footnote{Data taken from $https://lab24.ilsole24ore.com/coronavirus/ \;$.} and develop some previsions. Before getting into the specific details of the discussion, we give a short resume of the mathematical tools we will employ.\\

\begin{figure}[h]
	\centering
	\includegraphics[width=0.6\linewidth]{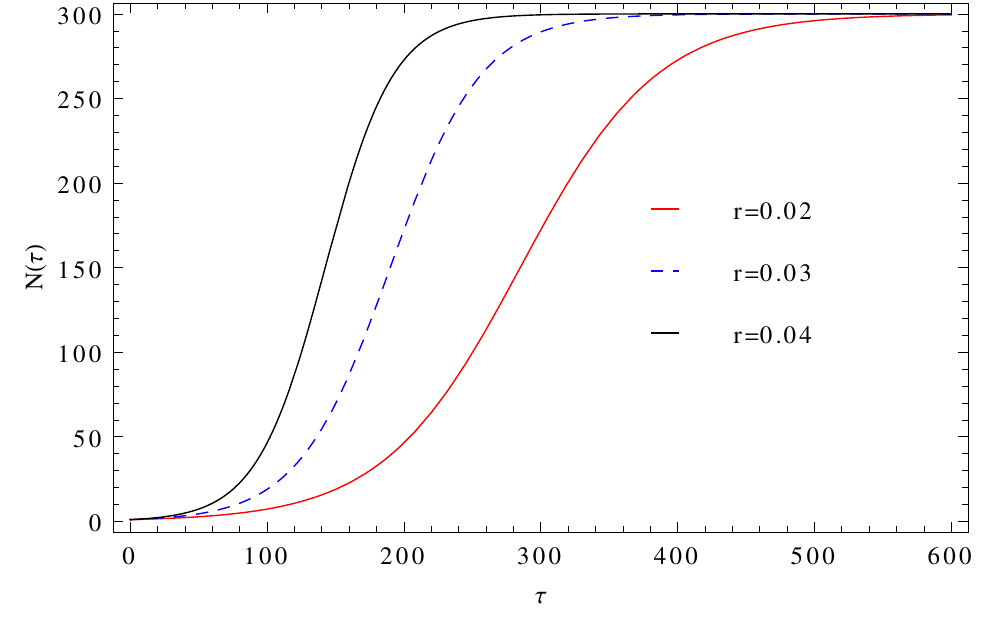}
	\caption{Growth of infected individual vs. $\tau$ for $K=300$ and different values of the growth rate $r$.}
	\label{fig:codfig1}
\end{figure}

The logistic function (Fig. \ref{fig:codfig1}) describing the evolution of a given population $N(\tau)$ is
\begin{equation}\label{nt}
N(\tau)=N_0\dfrac{e^{r\tau}}{1+\dfrac{N_0}{K}(e^{r\tau}-1)},
\end{equation}
 where $\tau$ is the time, measured in some units to be specified. Furthermore $r$ and $K$ accounts for the growth rate and the \textbf{carrying capacity} respectively. Regarding bacteria or viruses, $K$ represents the maximum number of individuals which can be infected. In absence of an ad hoc developed infectious model it cannot be easily specified (see the concluding section for further comments), it (and $r$ as well) will be inferred from the ``experimental data".\\
 
 The logistic equation provides a good model for the exponential growth and it does include the so called saturation mechanism leading to the equilibrium, where the population saturates. It is the solution of the first order non- linear equation \cite{Weisstei}
 \begin{equation}\label{key}
 \partial_\tau N=rN\left(1-\dfrac{N}{K} \right) 
 \end{equation}
which states that the growth rate is not simply $r$  but $r\left( 1-\frac{N}{K}\right) $. The logistic curve describes the total number of infections and its derivative yields the number of infections per unit time, namely 
\begin{equation}\label{key}
N'(\tau)=\frac{e^{rt} r\left( \dfrac{N_0}{1-\frac{N_0}{K}}\right) }{\left( 1+\dfrac{1}{K}\left( \dfrac{N_0}{1-\frac{N_0}{K}}\right)e^{r t}\right)^2 }=
\frac{e^{rt} rN_0(K-N_0) }{K\left( 1+\dfrac{N_0}{K}(e^{r t}-1)\right)^2 }.
\end{equation}
 It is a bell shaped curve and it is known as the \textbf{Hubbert curve}, initially exploited as an approximation of the production rate of a resource over time \cite{Deffeyes}. The location of the peak depends on growth rate and carrying capacity according to the identity 
\begin{equation}\label{key}
\tau^ *=\ln\left(\sqrt[r]{\dfrac{K}{N_0} -1}\right) .
\end{equation}
At this time we find for the infected rate
 \begin{equation}\label{key}
 N'(\tau^ *)=\dfrac{r K}{4}
 \end{equation}
corresponding to a total number of infected
\begin{equation}\label{key}
N(\tau^*)=\dfrac{K}{2}.
\end{equation}
The figures reported below yields a better idea of what has been worded so far.\\

In Fig. \ref{fig:codfig1} we have shown the growth of infected individual $N(\tau)$ vs. the number of days ($\tau=0$ has been arbitrarily chosen) for different values of the growth rate $r$ and for the same value of the carrying capacity $K$.\\

\begin{figure}[h]
	\centering
	\includegraphics[width=0.6\linewidth]{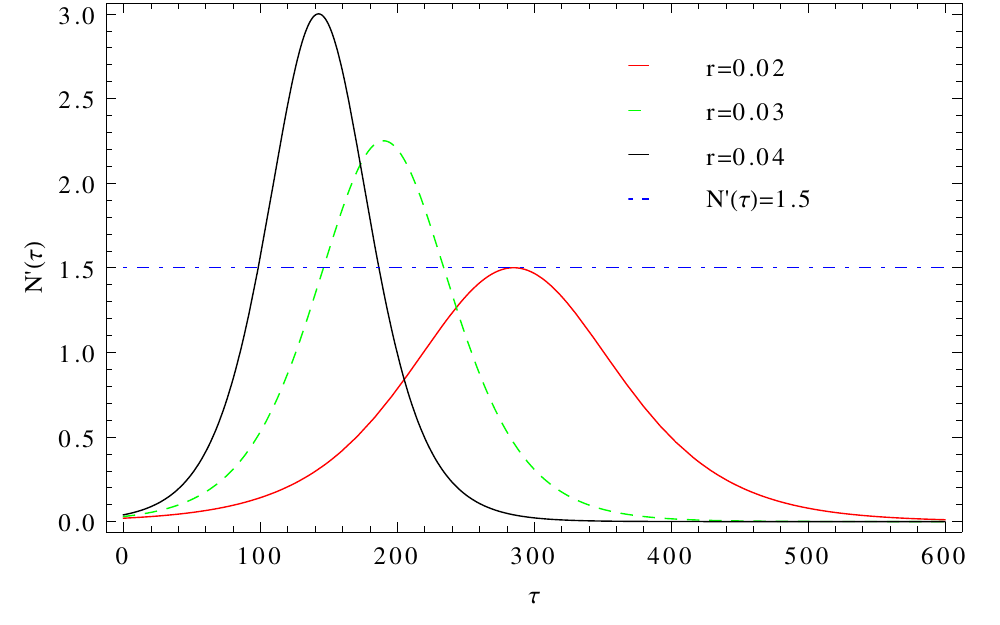}
	\caption{Hubbert curve vs. $\tau$ for $K=300$ and different values of the growth rate $r$ and a hypothetical threshold rate.}
	\label{fig:codfig2}
\end{figure}

In Fig. \ref{fig:codfig2} we have reported the Hubbert curve vs. $\tau$, for different values of $r$ and the same $K$. It is worth noting that, with decreasing $r$, the peak shifts towards larger $\tau$ values. We have fixed a threshold value the maximum rate that a health structure may care ($N'(\tau^*)$), the number of patients which cannot be treated is given by the total number of infected above this threshold ($\overline{N}$), namely 
\begin{equation}\label{key}
\begin{split}
& \Delta= N(\tau_+)-N(\tau_-), \\[1.2ex] 
& \tau_\pm=\dfrac{1}{r}\ln \left\lbrace \dfrac{K-N_0}{2N_0\overline{N}}\left[ r\;K -2\overline{N}\pm \sqrt{r\;K\;\left( r\;K-4\overline{N}\right) } \right]  \right\rbrace 
\end{split}
\end{equation}
where $\tau_\pm$ are the times in which the Hubbert and threshold curves intercept. \\

\begin{figure}[h]
	\centering
	\includegraphics[width=0.6\linewidth]{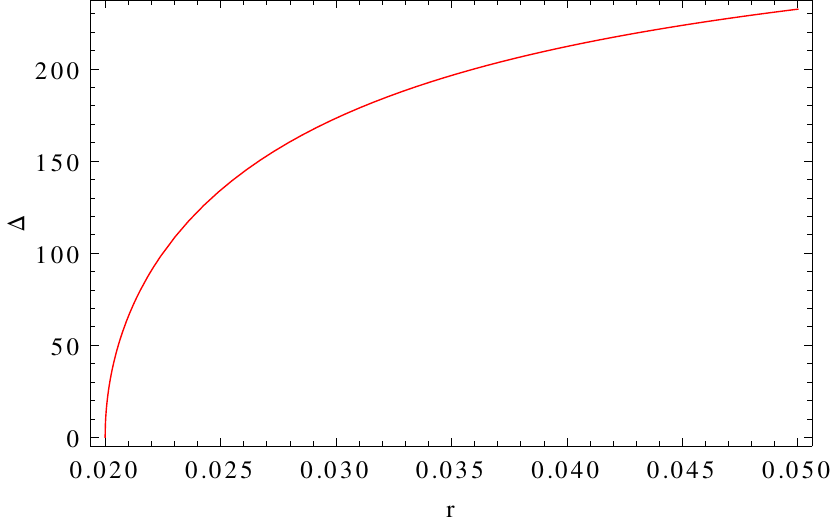}
	\caption{Number of patients potentially untreated vs. $r$ for $K=300$, $\overline{N}=1.5$, $N_0=1$.}
	\label{fig:codfig3}
\end{figure}

In Fig. \ref{fig:codfig3} we have reported the number $\Delta$ of patients potentially untreated vs $r$. \\

This is the minimum Mathematica we need to start our discussion. Further information will be added once needed.

\newpage

\section{Logistic Model and Covid-19 Data in Italy}

In this section we apply the formalism of the previous section to the analysis of the behavior of the viral infection in Italy.\\

\begin{figure}[h]
	\centering
	\includegraphics[width=.6\linewidth]{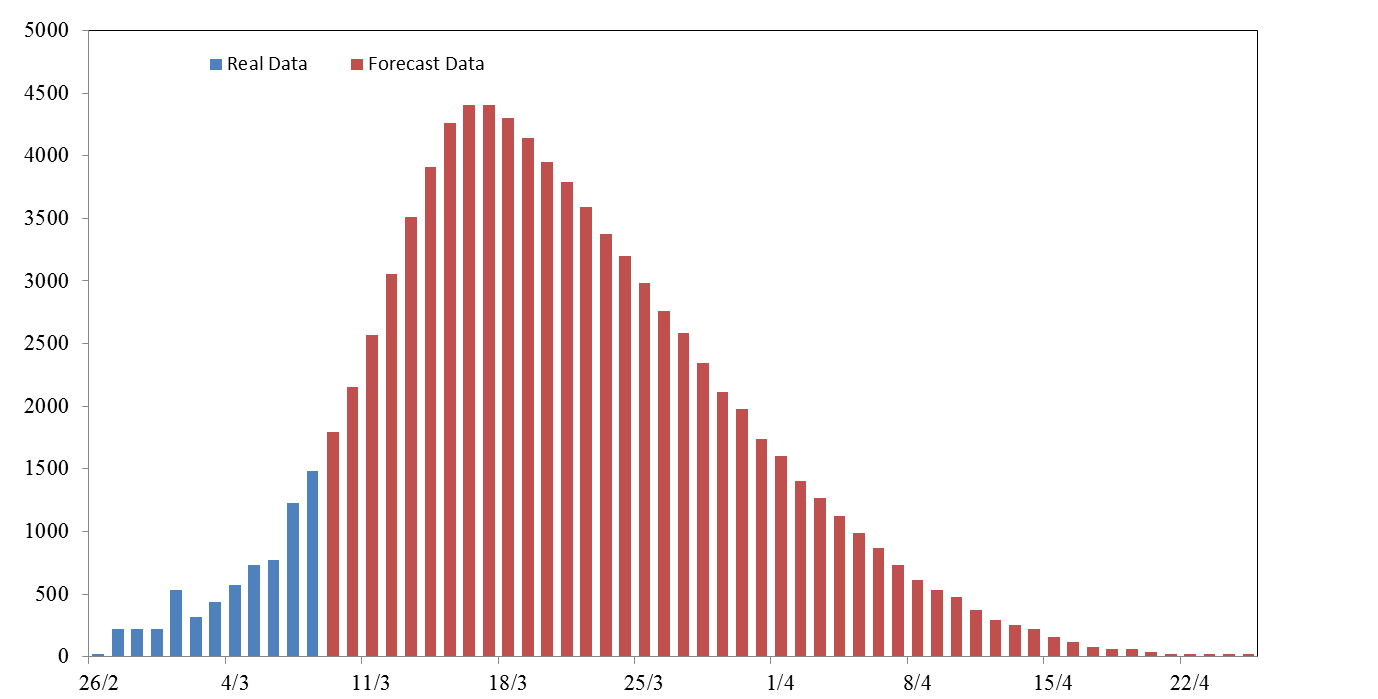}
	\caption{Real and foreseen number of infections per day from the beginning of the epidemic in Italy.}
	\label{fig:codfig4}
\end{figure}

The starting point is the number of infected individuals per day, provided by the Italian Ministry of Health\footnote{Data from $https://www.ilsole24ore.com/art/coronavirus-governo-stima-92mila-contagi-picco-18-marzo-ADfgS9C$. }. The data are reported in Fig. \ref{fig:codfig4}, with the inclusion of the forecasting of the 
future trend, which foresees the saturation by the end of April.\\
 
\begin{figure}[h]
		\begin{subfigure}{0.49\textwidth}
		\includegraphics[width=1.1\linewidth]{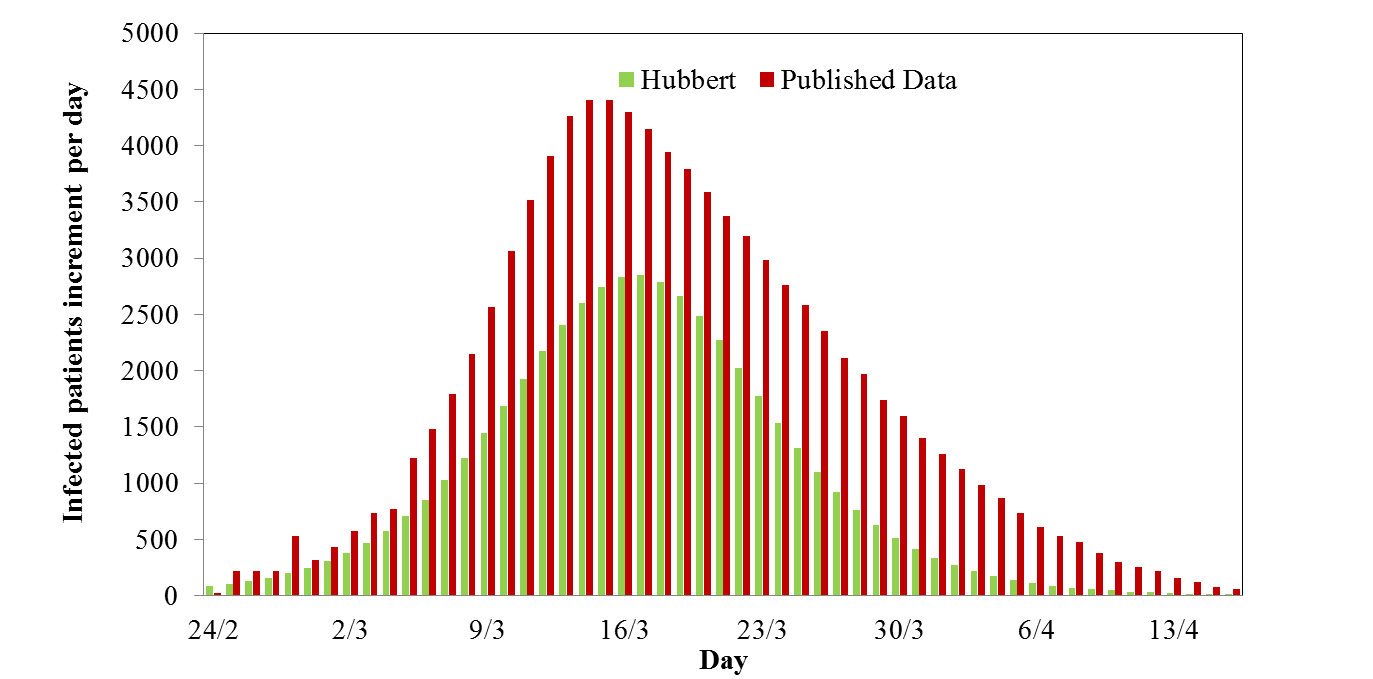}
		\caption{Comparison between the present model (green istogram) and Fig. \ref{fig:codfig4} (red).}
			\label{fig:codfig5a}
	\end{subfigure}\;
	\begin{subfigure}{0.49\textwidth}
	\includegraphics[width=1.\linewidth]{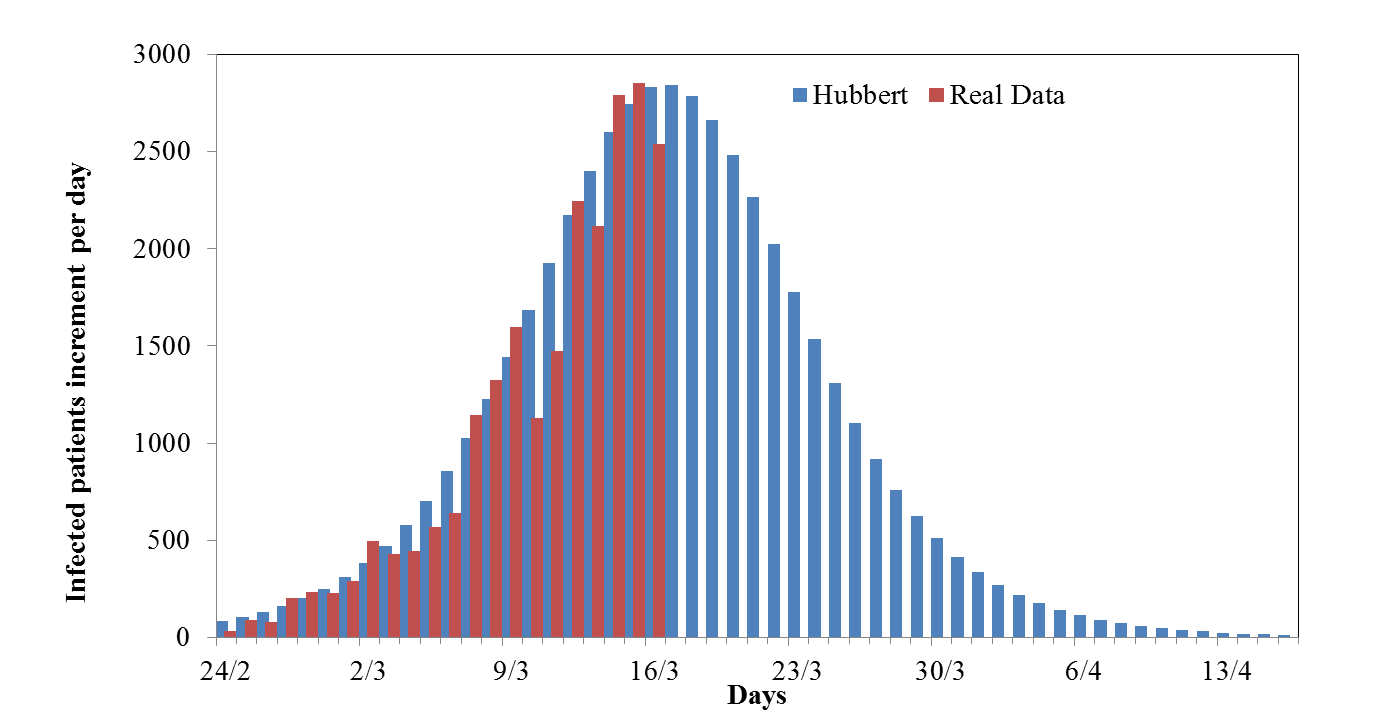}
	\caption{Comparison between Hubbert function and real data including $23$ days of data.}
		\label{fig:codfig5b}
\end{subfigure}
\caption{Real and foreseen numbers of infection per day from the beginning of the epidemic in Italy.}
	\label{fig:codfig5}
\end{figure}

 It is evident that the plot in Fig. \ref{fig:codfig4} is essentially the Hubbert curve and can be derived by a corresponding Logistic curve, as it shown in Figs. \ref{fig:codfig5} where we have reported the number of infected/day. In \ref{fig:codfig5a} we make the comparison between the present model and that in Fig. \ref{fig:codfig4}. To make a correct comparison we have fitted the logistic function parameter using the data of the first $12$ days. In Fig. \ref{fig:codfig5b} we have extended the fit to $23$ days (red istogram) and included the prevision from the our model (blue istogram).\\
 
%

\begin{figure}[h]
		\centering
	\begin{subfigure}[c]{0.48\textwidth}
	\includegraphics[width=1.1\linewidth]{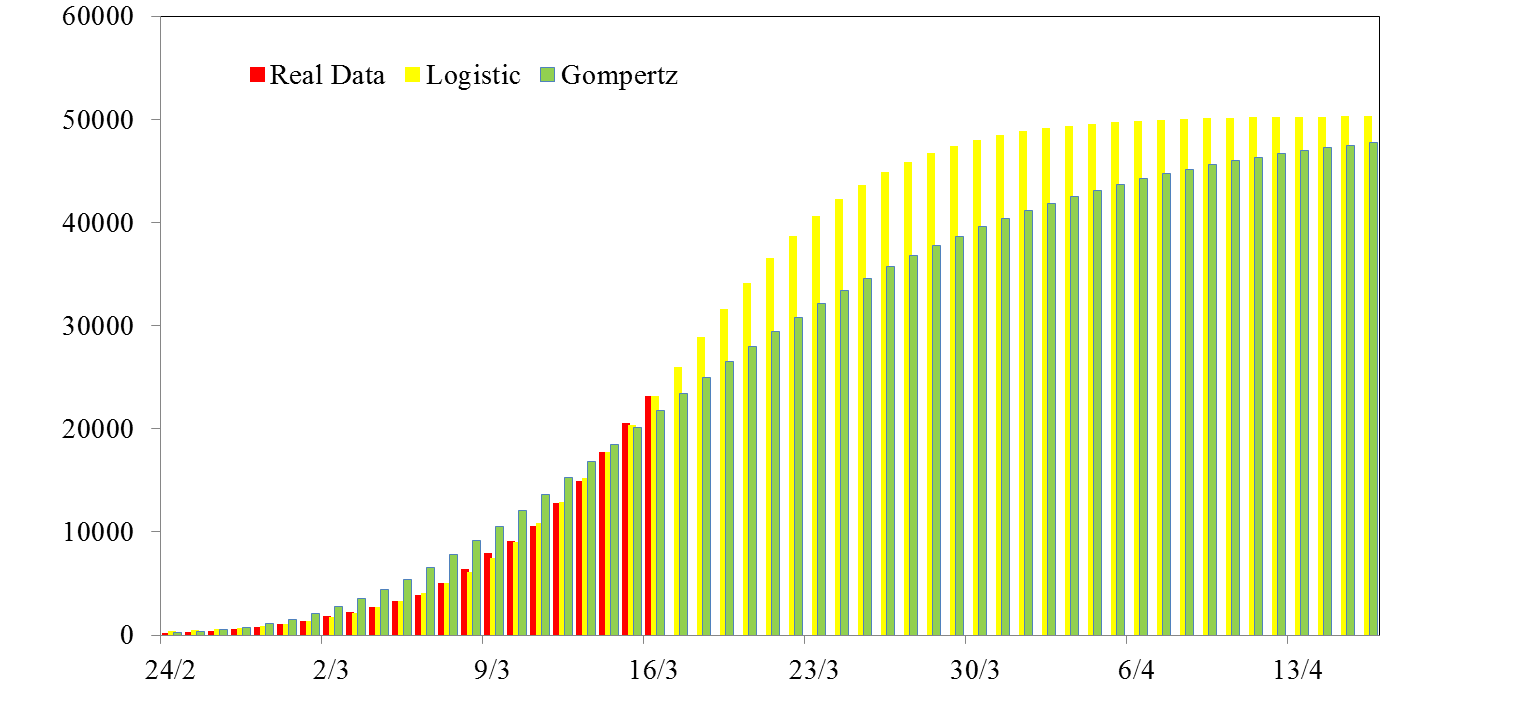}
	\caption{Evolution hystogram including data till March 16.}
	\label{fig:codfig7Emaa}
		\end{subfigure}\;
		\begin{subfigure}[c]{0.48\textwidth}
		\includegraphics[width=.9\linewidth]{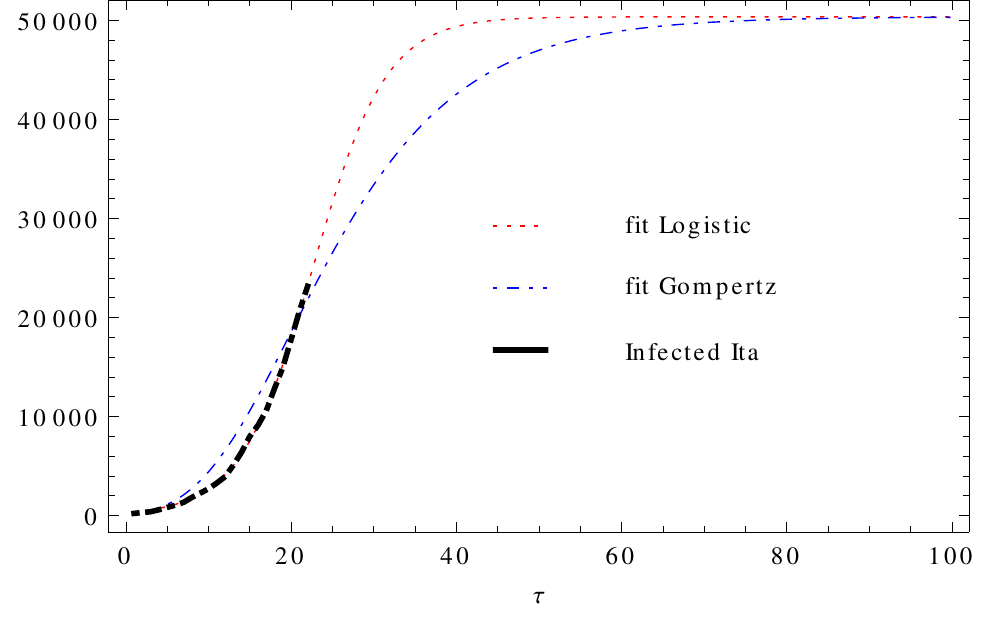}
		\caption{Evolution vs. $\tau$.}
		\label{fig:codfig7my}
	\end{subfigure}
\caption{Evolution of the infection (number of positive individuals to Covid 19 in Italy) vs. the number of days during which the morbidity has developed (for the Gompertz case see section 3).}
\label{fig:codfig7Ema}
\end{figure}

\noindent The fit of the logistic function with the available data yields the plot in Fig. \ref{fig:codfig7Ema}. It displays the evolution of the infection (number of positive individuals to Covid 19) vs. the number of days during which the morbidity has developed. 
\begin{itemize}
	\item [$\star$] The “zero” day, coincides with that in which the number of positive cases has  been started to be officially communicated. 
\item [$\star$] The green dotted curve represents the evolution of the infected according to the official data, while he red continuous is the logistic fit. The agreement looks satisfactory.
\item [$\star$] The best fit procedure yields 
\begin{equation}\label{N0rK}
N_0=294,  \qquad \qquad r=0.2264, \qquad \qquad K=50346.
\end{equation}
\end{itemize} 

\begin{figure}[h]
	\centering
	\includegraphics[width=0.6\linewidth]{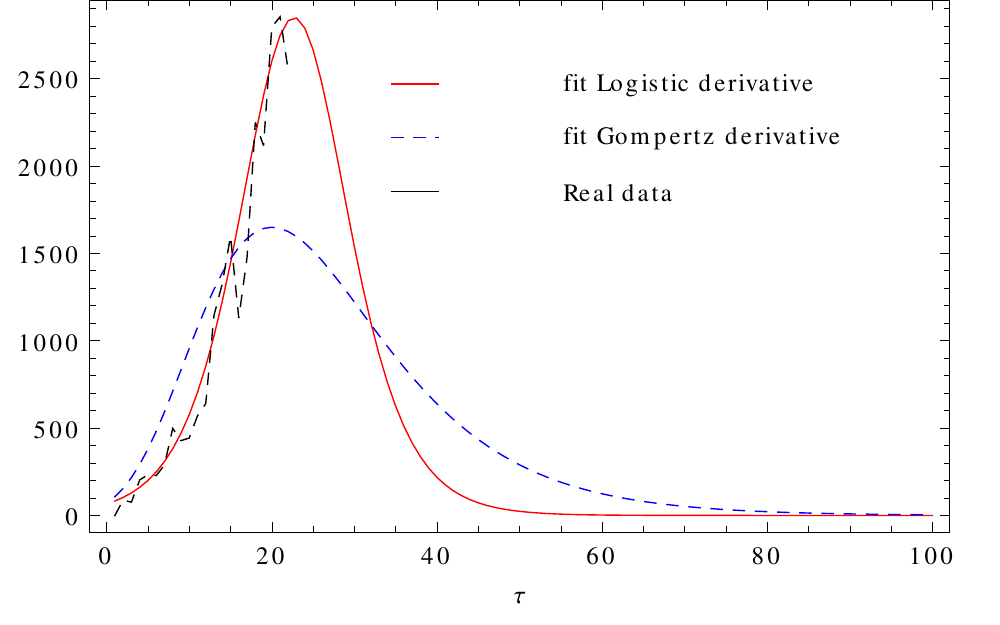}
	\caption{Comparison through Hubbert and Gompertz curves representing the number of Covid-19 positive per day, obtained from the fitted equations, and daily increment from the registered data (for the Gompertz case see section 3).}
	\label{fig:codfig6my}
\end{figure}

The Hubbert curve, inferred from the data in Figs. \ref{fig:codfig7Ema} is shown in Fig. \ref{fig:codfig6my} in which the continuous and dashed curves are just the analytical derivatives of the logistic functions, while the (noisy) black dashed curve is the increment of the data from public health ministry.\\

\begin{figure}[h]
	\centering
	\includegraphics[width=0.6\linewidth]{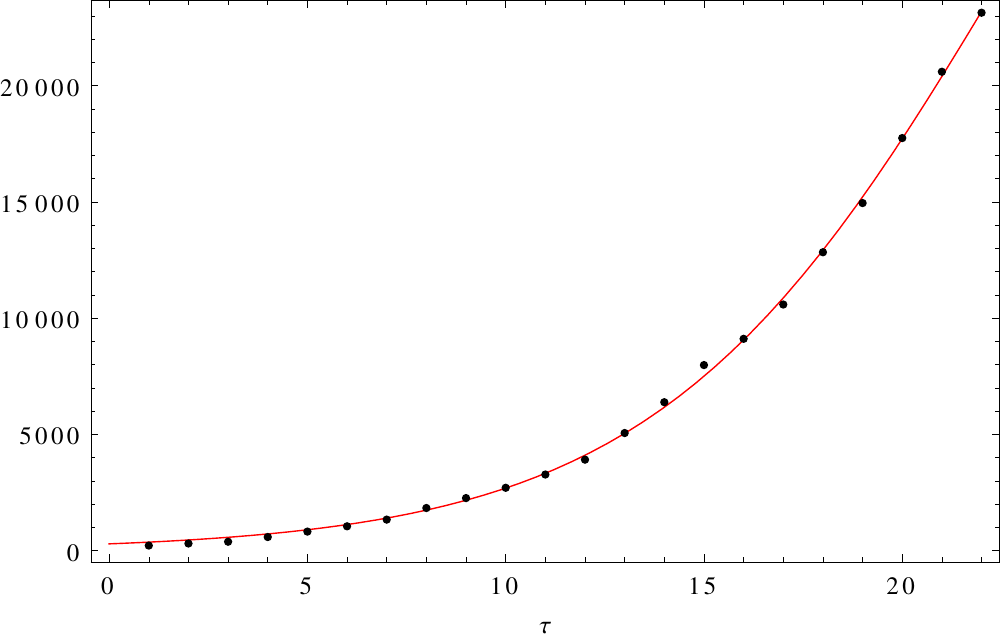}
	\caption{Comparison between empirical number of infected italian individuals in the period February 24 -- March 16, 2020, (black dotted) and Logistic equation \eqref{nt} for $N_0=294$, $r=0.2264$, $K=50346$.}
	\label{fig:codfig8my}
\end{figure}
The conclusions we may draw are that, if we trust this model, the peak of the number of infected per day should be reached in the next few days and that a substantive decline could be experienced at the end of April. \\

\textit{It should be noted that being the baricenter of the data shifted towards those of ``Regione Lombardia", the extension of the previous conclusions to the rest of the Nation is not fully justified. In the next days we will presumably observe the impact of the lockdown restriction and have a better picture of its role on parameters like the growth rate $r$.}\\

For further comments see the concluding section.\\

The number of positive cases around the peak is close to $3,000$, if $10\%$ needs intensive cares, about $300$ patient/per day are supposed to be allocated in health structures with limited resources. Further comments on these points will be given in the forthcoming section. 

\section{Final Comments}

The model we have exploited is a simplification in many sense. It does not include all the elements of the $SIR$ epidemic models of infections \cite{Brauer}. We considered infected only and we did not include the evolution of healed and mortality.\\

The logistic model seems to work in a reasonable way and, if further refined, may become a  predictive tool. We must however underline that a sigmoid evolution curve and a bell shaped Hubbert form can be derived within different analytical frameworks. The most attractive alternative is the so called Gompertz model \cite{Tjorve}, based on the evolution curve
\begin{equation}\label{eqG}
G(\tau)=\alpha e^{-\beta\; e^{-\gamma \tau}}.
\end{equation}
The analytical dependence on time is interesting for our purposes because it displays an exponential term, which simultaneously includes the growth rate and the saturation. Regarding the meaning of the parameters we note that
\begin{equation}\label{key}
G_0=G(0)=\alpha e^{-\beta}, \qquad \qquad \lim\limits_{\tau\rightarrow\infty}=\alpha .
\end{equation}
We recognize $\alpha$ as the carrying capacity so that
\begin{equation}\label{key}
\alpha=K, \qquad \qquad \beta=-\ln\left(\dfrac{G_0}{K} \right) 
\end{equation}
and eventually, with $N_0=G_0$, we can write Eq. \eqref{eqG} as
\begin{equation}\label{key}
N(\tau)=G(\tau)=K\left(\dfrac{N_0}{K} \right)^{e^{-\gamma\tau}} 
\end{equation}
If we keep $N_0$ and $K$ the same as those predicted by the Logistic model and derive $\gamma$ from the data, we can evaluate the corresponding Hubbert curve, as reported in Figs. \ref{fig:codfig7Ema} in which we have made a comparison with the logistic case.\\

The most remarkable difference is that the Gompertz-Hubbert curve is more asymmetric and wider than the Logistic counterpart, this behavior seems more similar to the case reported in Fig. \ref{fig:codfig4} (which is also the result of a computational model). The drawback is that the two models predict different values of the peaks which, by using parametrs from Eq. \eqref{N0rK}, in the case of the logistic is 
\begin{equation}\label{key}
 max(N'(\tau))=2850 \qquad \qquad reached\; at\; \tau^*= 23, 
\end{equation}
and of the Gompertz
\begin{equation}\label{key}
G'(\tau)=\alpha\;\beta\;\gamma\;
 e^{-\gamma \tau} e^{-\beta e^{-\gamma \tau}}=-\gamma \ln\left(\dfrac{N_0}{K} \right) e^{-\gamma \tau}G(\tau)
\end{equation}
 is
\begin{equation}\label{key}
 max(G'(\tau))=1648 \quad \quad reached\; at\; \tau_G^*= 18, \qquad \qquad  \tau_G^*=\dfrac{1}{\gamma}\ln\left(\ln\left(\dfrac{K}{N_0} \right)  \right)
 \end{equation}
 where
 \begin{equation}
 \begin{split}
& G(\tau_G^*)=K\left(\dfrac{N_0}{K} \right)^{\left(\ln\left( \frac{K}{N_0}\right)  \right)^{-1} }, \\
&  G'(\tau_G^*)=-K\;\gamma\;\ln\left( \dfrac{N_0}{K}\right)\;\left(\dfrac{N_0}{K} \right)^{\left(\ln\left( \frac{K}{N_0}\right)  \right)^{-1} } \left( \ln\left( \frac{K}{N_0}\right)\right) ^{-1}\;.
\end{split}
\end{equation}

The conclusions we can draw confirm essentially what we have underscored in the previous section, with the caveat that the data are referring to the whole national territory, in which the infection is proceeding in a non homogeneous way. \\

\begin{figure}[h]
	\centering
	\includegraphics[width=0.8\linewidth]{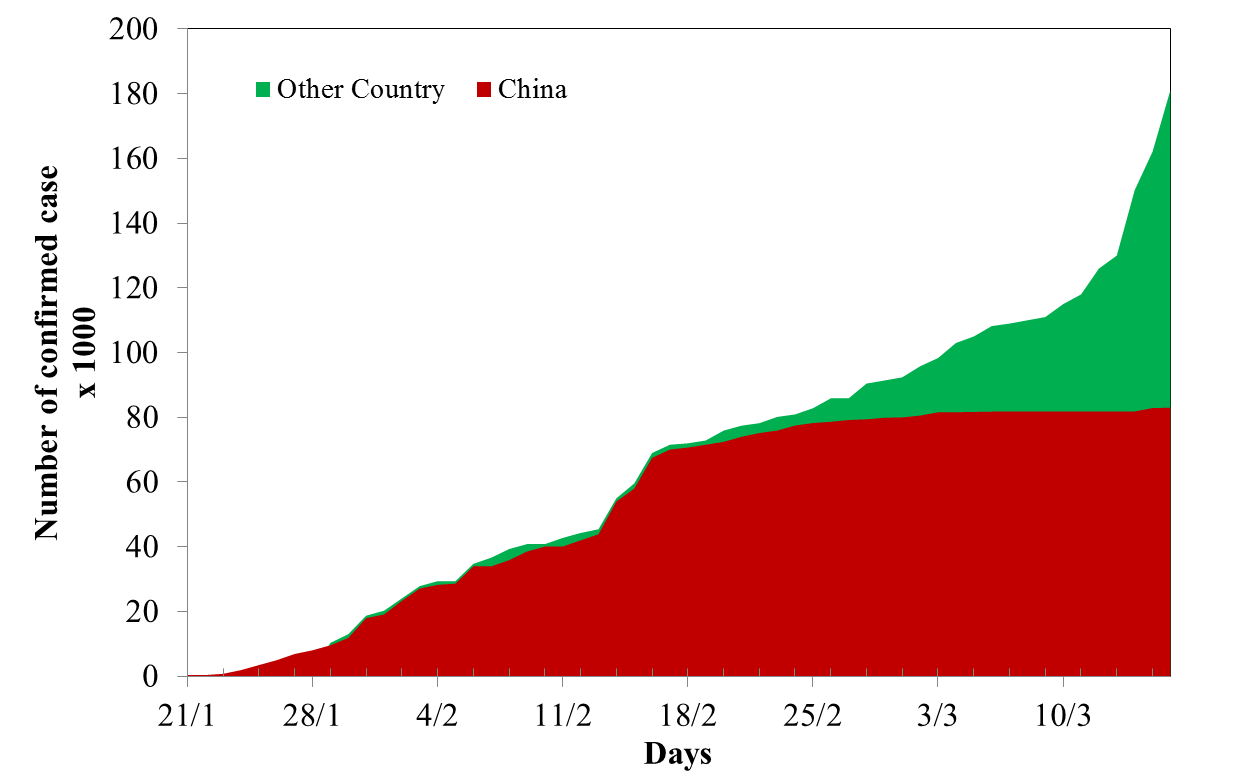}
	\caption{World wide evolution of the disease.
		The pedestal of the evolution is a logistic type curve, representing the spread in China. After a certain time lag the epidemy has propagated in the rest of the world, starting a second logistic behavior (data from Johns Hopkins CSSE,WHO, CDC, NHC and DINGXIANGYUAN).}
	\label{fig:codfig8}
\end{figure}

Our analysis is strongly influenced by the growth of the disease in Lombardia (and more in general in the north of Italy). The evolution in the national territory might qualitatively reproduce that reported in Fig. \ref{fig:codfig8}, which refers to the world- wide landscape. 
The growth of the disease in other regions, owing to the lockdown, may follow a different pattern. We can foresee a kind of saturation and then a resurgence at lower increasing rate, if the government prescription are duly observed.
In qualitative  terms we may observe the same bi-logistic behavior \cite{Meyer}, with China replacing the north of Italy and the superimposed plot yielding the diffusion in the rest of Italy. In a forthcoming note we will analyze the evolution per individual region.\\

We must finally stress that the spatial diffusion of the disease, not contained in the present analysis, requires the use of models including non-linear Fisher type evolution equations \cite{Fisher2,Fisher}, which will be touched in forthcoming studies.

\textbf{Acknowledgements}\\

The work of Dr. S. Licciardi was supported by an Enea Research Center individual fellowship.\\
\indent The Authors express their sincere appreciation to Dr. Ada A. Dattoli for her help in understanding the biological basis of the infection.\\





\begin{thebibliography}{}

\bibitem{Allen} Allen, R.A, Waclaw, B.
Bacterial growth: a statistical physicist’s guide,
\textit{Rep Prog Phys.}, 82(1): 016601, 2019. 

\bibitem{Fenchel} Fenchel, T., King, G.M., Blackburn, T.H. Bacterial Biogeochemistry: The Ecophysiology of Mineral Cycling, Elsevier; 1998. 

\bibitem{Flint} Flint, H.J, Scott, K.P., Louis, P., Duncan, S.H. The role of the gut microbiota in nutrition and health, \textit{Nat Rev Gastro Hepat.}, 9: pp. 577--589, 2012.

\bibitem{Sevious} Sevious, P., Halkjaer Nielsen, P., Microbial ecology of activated sludge, IWA Publishing; 2010. 

\bibitem{DGDO} Dattoli, G., Guiot, C., Delsanto, P.P., Ottaviani, P.L., Pagnutti, S., Deisboeck, T.S.,
\textit{J. Theor. Biol.}, vol. 256, 3, pp. 305--310, 2009.

\bibitem{DFel} Dattoli, G., Di Palma, E., Pagnutti, S., Sabia, E., Free Electron coherent sources: From microwave to X-rays, \textit{Phys. Rep.}, 739, pp. 1--52, 2018.

\bibitem{Cramer} Cramer, J.S., The origin of Logistic Regression, TI 2002 119/4, Tinbergen Institute Discussion Paper.

\bibitem{Tjorve} Tjorve, K.M.C., Tjorve, E.,
The use of Gompertz models in growth
analyses, and new Gompertz-model
approach: An addition to the Unified-Richards family, \textit{PLOS one}, 2017, doi.org/10.1371/journal.pone.0178691.

\bibitem{Weisstei} Weisstein, E.W., Logistic Equation, From MathWorld--A Wolfram Web Resource, http://mathworld.wolfram.com/LogisticEquation.html .

\bibitem{Deffeyes} Deffeyes, K.S., Hubbert's Peak: The Impending World Oil Shortage, Published by: Princeton University Press, 2008.

\bibitem{Brauer} Brauer, F., Castillo-Chávez, C., Mathematical Models in Population Biology and Epidemiology, NY: Springer, 2001

\bibitem{Meyer} Meyer, P.S:, Bi-Logistic Growth, Technological Forecasting and Social Change 47: pp. 89--102, 1994.

\bibitem{Fisher2} Fisher, R.A., The Wave of Advance of Advantageous Genes, \textit{Annals of Eugenics}, 1937.

\bibitem{Fisher} Dattoli, G., Di Palma, E., Sabia, E., Licciardi, S.,Quasi Exact Solution of the Fisher Equation, \textit{Appl. Math.}, vol. 4, 8A, pp. 7--12, 2013. 

%
%
%
%
%
%
%
%
%
%
%
%
%
%
%
%
%
%
%
%
%
 

\end{thebibliography}
\end{document}